\documentclass[12pt,preprint]{aastex}

\newcommand{\cerenkov}{{\v C}erenkov}
\newcommand{\sgra}{Sgr A*}
\newcommand{\grs}{$\gamma$-rays}

\newcommand{\etal}{\emph{et al.}}

\def\3EG{{3EG J1746-2851}}
\def\p0{{$\pi^0$}}
\def\1018{{$10^{18}$}}

\newcommand{\be}{\begin{equation}}
\newcommand{\ee}{\end{equation}}
\newcommand{\bea}{\begin{eqnarray}}
\newcommand{\eea}{\end{eqnarray}}
\newcommand{\bmu}{\begin{multline}}
\newcommand{\emu}{\end{multline}}

\def\simlt{\lower.5ex\hbox{$\; \buildrel < \over \sim \;$}}
\def\simgt{\lower.5ex\hbox{$\; \buildrel > \over \sim \;$}}

\def\gcm3{{\rm\,g\,cm^{-3}}}
\def\ncm3{{\rm\,cm^{-3}}}

\def\>{$>$}
\def\<{$<$}

\shorttitle{Synchrotron from Secondary Leptons near Sgr A*}
\shortauthors{Crocker \etal}

\begin{document}

\title{Radio Synchrotron Emission from Secondary Leptons in the Vicinity of Sgr A*}


\author{Roland M. Crocker\altaffilmark{1}, David Jones\altaffilmark{1,2}, David R. Ballantyne\altaffilmark{3},
and Fulvio Melia\altaffilmark{3,4,5}}

\altaffiltext{1}{School of Chemistry and Physics, The University of
  Adelaide, Adelaide, South Australia, 5005 Australia;
  roland.crocker,david.jones@adelaide.edu.au}
\altaffiltext{2}{Australia Telescope National Facility, Marsfield,2122, Australia}
\altaffiltext{3}{Department of Physics, The University of Arizona, 1118 East 4th
  Street, Tucson, AZ 85721; drb, melia@physics.arizona.edu}
\altaffiltext{4}{Steward Observatory, The University of Arizona, 933
  N. Cherry Avenue, Tucson, AZ 85721}
\altaffiltext{5}{Sir Thomas Lyle Fellow and Miegunyah Fellow.}

\begin{abstract}

A point-like source of $\sim$TeV \grs\ has recently been seen towards the
Galactic center by
HESS and other air \ \cerenkov\ telescopes.
In recent work \citep{Ballantyne2007}, we demonstrated that these \grs\
can be attributed to high-energy protons
that (i) are accelerated
close to the event horizon
of the central black hole, \sgra , (ii) diffuse out to
$\sim$ pc scales, and (iii) finally 
interact to produce \grs.
The same hadronic collision processes will necessarily lead to the creation of 
electrons and positrons.
Here
we calculate the synchrotron emissivity of these
secondary leptons in the same magnetic field configuration
through which the initiating protons have been propagated in our model.
We compare this emission with the observed $\sim$ GHz radio spectrum of the
inner few pc region which we have assembled from
archival data and new measurements
we have made with the
Australia Telescope Compact Array.
We find that our model predicts secondary synchrotron emission
with a steep slope consistent with the observations but with an overall normalization
that is too large by a factor of $\sim$ 2.
If we further constrain our theoretical $\gamma$-ray curve
to obey the implict EGRET upper limit on emission from this region
we predict radio emission that is consistent with
observations, i.e., the hadronic model
of gamma ray emission can, simultaneously and 
without fine-tuning, also explain essentially all the diffuse radio emission 
detected from the inner few pc of the Galaxy.
\end{abstract}


\section{Introduction}
\label{sect:Introduction}
The Galactic center (GC), at an assumed distance of 7.9 kpc, 
is a complex region containing
many compact and
diffuse high-energy astrophysical sources
 (see \citealt{Melia2007} for a
recent review).
The dynamics of the central few lightyears is dominated by what is
believed to be a supermassive black hole, Sagittarius A* (Sgr A*).

The GC has been identified as a point-like source of TeV \grs\ by
a number of air
\cerenkov\ telescopes.
In our work
we concentrate exclusively on the
data from
HESS \citep{Aharonian2004,Aharonian2006b}
as these provide the best constraints on the source
properties.
The GC signal 
is coincident within $\sim 30\arcsec$ of \sgra, though with a
centroid displaced roughly $7\arcsec$ ($\sim 0.4$~pc) to the East of
the GC \citep{Aharonian2004,Aharonian2006b}.
The
spectrum is a pure power law with photon index
$2.25\pm0.10$, and the total flux above 1~TeV is $(1.87 \pm 0.30)
\times 10^{-8}$~m$^{-2}$~s$^{-1}$ \citep{Aharonian2006b}.

In a recent paper \citep{Ballantyne2007}, we examined a hadronic scenario
for the creation of the $\sim$TeV \grs:
a power-law population of protons 
is accelerated close to Sgr A*, perhaps
through the process of stochastic acceleration on the turbulent magnetic fields expected to pervade this inner
region.
These then
diffuse outwards 
through the 
strong magnetic fields of the largely evacuated stellar-wind region
surrounding Sgr A* before, finally,
colliding with gas in the high molecular density clumps that make up the circumnuclear disk (CND).
Such collisions (assumed to be exclusively proton on proton for simplicity) generate
neutral mesons that then decay into the \grs \ which, by construction, we detect here at Earth.

In the current work we self-consistently calculate the synchrotron emissivity of the
secondary leptons that must also be produced
in our scenario. 
These come from the decay of the charged mesons created
in the same p-p collisions that we posit explain
the $\gamma$-ray emission -- though, in constrast
to the case for $\gamma$-ray emission, we find that, because of the extremely strong
magnetic fields there (and despite the relatively low ambient densities)
radio emission from the wind region is  more important than that from the CND.
Finally, we compare our predictions 
for radio emission with archival and new radio data on the region we have assembled.

\section{Proton Propagation}
\label{sect:Calculations}

We firstly summarize the results obtained by \citet{Ballantyne2007}.
In our calculation we employed a realistic model
of the density of matter through the GC arrived at by \citet{Rockefeller2004}.
These authors computed the density distribution in
this region caused by the interactions of stellar winds from the
young stars surrounding \sgra .
In addition to the stellar wind gas, the
volume also contains a high-density torus of molecular gas
with an inner radius of
1.2~pc and a thickness of 1~pc representing the observed
CND.

The average
density in the model stellar-wind gas (taken
to be any region where $n_{\mathrm{H}} < 3\times 10^3$~cm$^{-3}$) is $\left <
n_{\mathrm{H}}^{\mathrm{sw}} \right > = 121$~cm$^{-3}$, while it is
$\left < n_{\mathrm{H}}^{\mathrm{mt}} \right > =$ 233,222~cm$^{-3}$ within
the model molecular torus. Taking $kT=1.3$~keV as the average temperature of the
stellar-wind gas \citep{Baganoff2003,Rockefeller2004}, 100~K for the temperature of
the molecular torus (\citealt{Rockefeller2004} and references therein), and
assuming equipartition, the average field intensity is $3$~mG
in the stellar-wind region and $0.35$~mG within the torus.
In any particular cell the 
magnetic field was assumed to be generated 
with an intensity that satisfies $|B| = 1.5 \times 10^{-5}$ Gauss $n_H/10^4$ cm$^{-3}$
for the CND region and $|B| = 2.6 \times 10^{-3}$ Gauss $n_H/10^2$ cm$^{-3}$
in the stellar-wind gas.
Our procedure for determining
a physical
description of the field direction at any position in the computational grid
is described in \citet{Ballantyne2007}.


Computational resources limited our modeling to proton
energies between $1$ and $100$~TeV. 
A total of 222,617 proton
trajectories was calculated with energies uniformly distributed in
this range.
These trajectories were split into $21$ energy bins
($\log (E/\mathrm{eV}) = [12.0-12.1], \ldots, [13.9-14.0]$).
From these data the steady state distribution of protons (given in cm$^{-3}$ eV $^{-1}$) in 
each of the 7500 computational cells of the CND and
the 7022 cells of the stellar-wind region
could be inferred.

As a proton random-walks its way 
through the the turbulent magnetic field of the GC it may collide
with a low-energy proton in the ambient medium and produce pions via the
reaction $p  +  p \rightarrow p  +  h  +  N_0 \pi^0  +  N_\pm \pi^\pm$. Here
$h$, denoting hadron, is a proton or, if charge exchange occurs, a neutron and
variable
and energy-dependent multiplicities of
neutral ($N_0$) and charged ($N_\pm$) pions are, in general, produced
but electric charge must be conserved overall.
(In our calculations we also account for the sub-dominant
contribution from charged and neutral kaons.)
Neutral pions will subsequently decay
into two photons and charged pions to electrons, positrons, and neutrinos.

In this Letter we adopt
the steady-state proton distributions within
each modeled clump of CND and stellar-wind region gas arrived at by \citet{Ballantyne2007}.
We then use the techniques outlined at length in
\citet{Crocker2007a} to calculate -- on the basis of (i) these steady state $p$
distributions, (ii) the ambient hydrogen number density, and (iii) clump
magnetic fields -- the steady state (processed) distributions of
secondary leptons within each clump and the resulting
synchrotron emission from each clump.

\section{Data}
\label{sect:Data}

To test our theoretical model we have obtained $\sim$GHz radio fluxes covering the region of the CND defined by a $2' \times 1'$
rectangle with sides parallel to Galactic longitude/latitude and with Sgr A* at the center.
Where possible,
we have chosen
interferometry data obtained with an array configuration that results in a beam smaller than this region
of interest to ameliorate the problem of confusion with nearby strong radio sources 
(Sgr A East in particular).
On the other hand, to ensure that the diffuse flux through this region that we are interested in --
present on scales up to the size of the region -- is not integrated out, we also demand
that the array configuration possess some antenna spacings that are suitably small.

For the reasons just given
we make use in our analysis of
medium resolution ($\sim43''$; \cite{LaRosa2000})
330 MHz VLA\footnote{The Very Large Array (VLA), as part of the National Radio Astronomy Observatory,
is operated
by Associated Universities, Inc., under cooperative agreement with the National Science Foundation.}
data, similar resolution SUMSS\footnote{The Molonglo Observatory Synthesis Telescope (MOST)
is operated by the University of Sydney} 843 MHz data, and serendipitous
ATCA\footnote{The Australia Telescope Compact Array (ATCA) is operated by the Australia Telescope National Facility,
CSIRO, as a National Research Facility.} observations of the GC at 1384 MHz and 2368 MHz.
These latter data were obtained with a somewhat lower resolution
($\sim 1'\times 2'$ at 1384 MHz, and $\sim 0.5'\times 1'$ at 2368 MHz.)
which is nevertheless sensitive to emission on the size scales of the CND.

For each radio image, 
we used a number of independent methods to obtain
the flux inside the region of interest (see  \cite{Jones2007} for details).
To obtain a conservative estimate on the error
in the flux determinations, the standard deviation between
all flux estimates at a particular frequency was taken to be the RMS error on the final flux quoted.
(Errors due to gain and to image noise make a negligible contribution).
The quoted central value for the flux at a given frequency is obtained
from smearing the beam to $2'\times1'$ resolution, and reading the peak flux 
which provides the most reliable determination.

Flux determinations are:
$46 \pm 12$ Jy at 330 MHz, $37 \pm 5$ Jy at 843 MHz, $20 \pm 6$ Jy at 1384 MHz, and
$11 \pm 2$ Jy at 2368 MHz.
These data are shown as the lower frequency data points in Figure \ref{fig:CNDRadioSpectrum} (b).
The data points at 843, 1384, and 2368 MHz define a non-thermal, power-law spectrum
in frequency $\propto \nu^{-1.2}$, indicating a steep, synchrotron-radiating
electron population (whether primary or secondary) of $\propto
E_e^{-3.4}$. 
The presence of non-thermal emission near Sgr A West was first discussed by
\citet{Gopal-Krishna1976} and was subsequently extensively investigated by
\citet{Ekers1983}. 
We note that the spectrum we determined for the region is steeper than that 
found by these latter authors though the overall normalization at 1384 MHz is compatible.
Our datum at 330 MHz falls below the $E_e^{-3.4}$ power law; this, however, is as expected
as Sgr A Complex radio emission
is strongly attenuated at around this frequency
and below because of
free-free absorption by thermal gas associated with the Sgr A West structure
as established by \citet{Pedlar1989}.
It is interesting to note that, according to these authors (see also \cite{Fatuzzo2003}), the 
spectral index pertaining to Sgr A East as a whole is approximately 1 with a corresponding average flux density 
at 1384 MHz of approximately 0.014 Jy/arcsec$^2$.
This is roughly half of the flux density we measure from the CND region.

At frequencies higher than 2368 MHz, 
we have obtained snapshot observations 
with ATCA
at 4800 and 8640 MHz. We measure 22 and 21 Jy at 4800 and 8640 MHz
over the same solid angle as for the lower frequency observations
and
compatible with the results obtained by \citet{Brown1981} at 4.9 GHz.  
Finally, in Figure \ref{fig:CNDRadioSpectrum} (b), 
we also show fluxes measured by 
\citet{Salter1988} at 84 and 230 GHz,
\citet{Sofue1986} at 43.25 GHz,
and \citet{Tsuboi1988} at 91 GHz.  
These latter measurements are not perfectly matched in solid angle to the region of concern and should be taken as indicative only.  
Overall, they show that the spectrum of the CND region has become dominated by thermal emission 
from Sgr A West at $\gtrsim$ 4 GHz and do not provide a strong constraint on our secondary synchrotron emission model.

\section{Results and Discussion}
\label{sect:Discussion}

From the modeling described in \citet{Ballantyne2007},
we have simulated data describing the steady-state proton distribution for
the clumps of matter
that make up the CND 
and the stellar-wind region
(each of varying density, magnetic field strength, and position).

Synchrotron radiation in the GHz range by secondary leptons in the mean magnetic field of the CND, 0.35 mG,
requires initiating parent $p$'s of energy $\sim 6 \times 10^{10}$ eV 
and only $\sim 2 \times 10^{10}$ eV in the 3 mG mean field of the wind region
(assuming a power-law $p$ spectrum of spectral index 2.3).
Such energies 
are unfortunately well below the 1 TeV threshold 
of our proton propagation modeling.
We assume, then, that the steady-state proton distribution
of any given computational cell is a pure power law (in momentum)
with such power law fitted to match our modeled 1-100 TeV proton spectra.

Despite our simulating more than $2 \times 10^5$ proton trajectories,
we must deal (in our power-law fitting) with 
empty energy bins in many clumps.
We address this issue by producing (on the basis of our modeled, {\it overall} clump spectra)
parameterizations of the TeV proton flux and fitted spectral index in terms of the
clump radial separation from Sgr A* and magnetic field strength.
With these parameterizations in hand we can determine, for any clump at given
radius and with given magnetic field, the {\it expected}
value for the $p$ flux in a given 10$^\mathrm{ th}$-decade energy bin.
For a zero-entry bin, we replace the zero with this expected value divided by 
the free parameter $X$.
We then use a MATHEMATICA routine to perform a $\chi^2$ fit in the parameter $X$ of
the overall CND TeV $\gamma$-ray spectrum to the HESS data
(the $\gamma$-ray emission by cells in the wind region only contributes at the $\sim$ 0.1\% level
and is ignored here).

\begin{figure}
\includegraphics[width=0.4\textwidth]{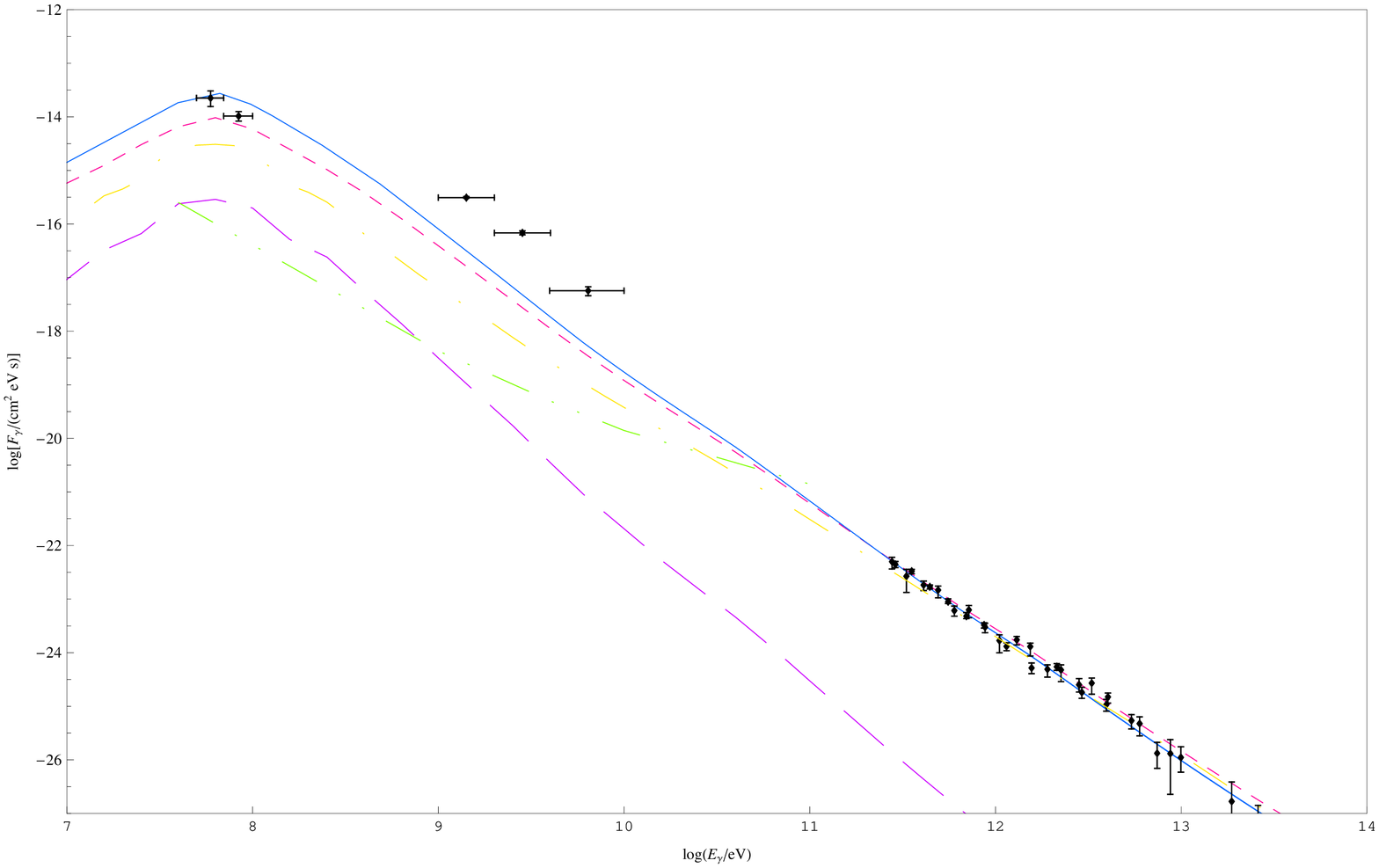}
\includegraphics[width=0.4\textwidth]{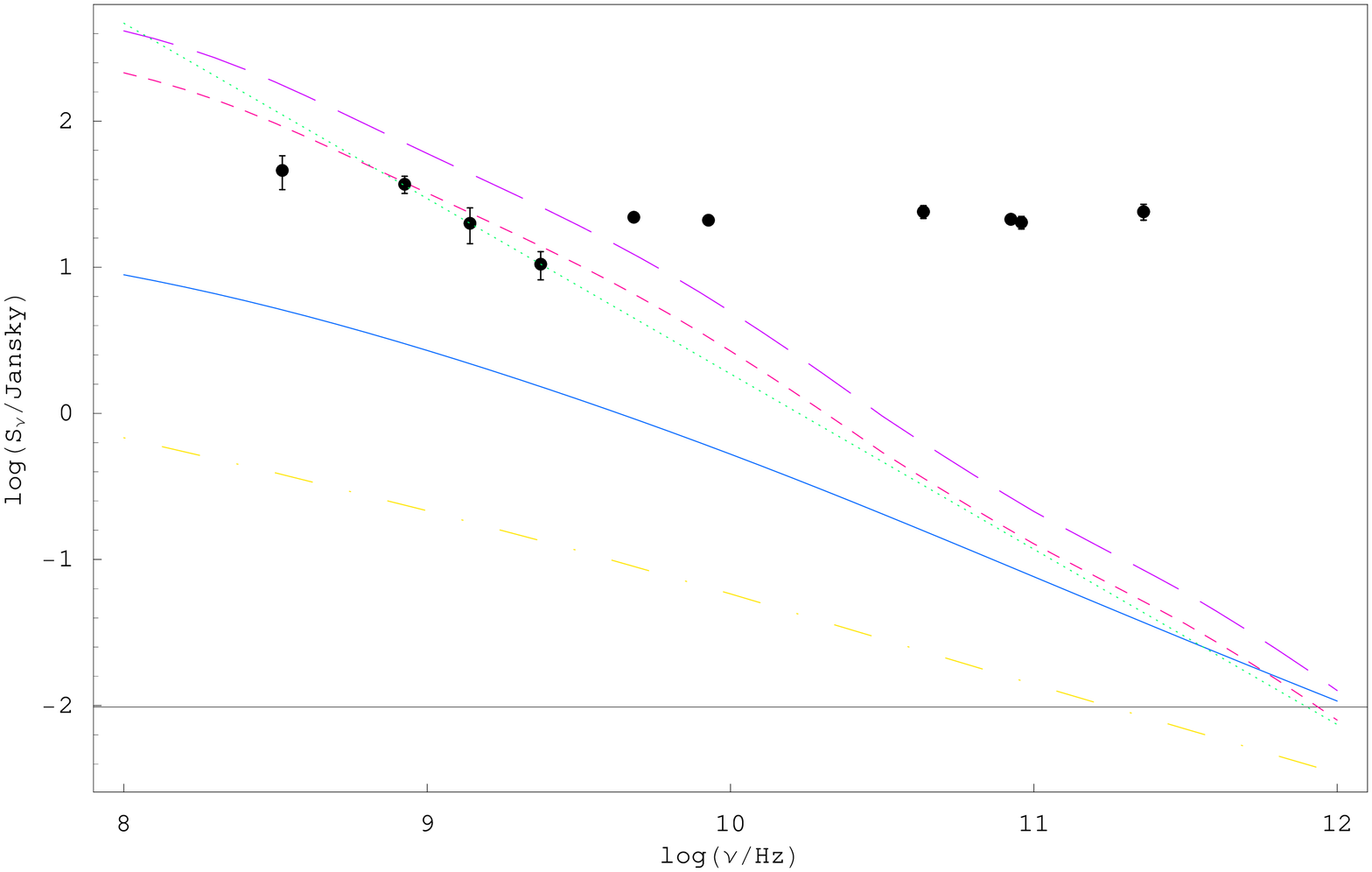}
\caption{{\bf a.} ({\bf left}) Observed $\gamma$-ray flux data points for the GC region with 
theoretical spectral curves.
The data points in the $\sim$ TeV range show the HESS source coincident with Sgr A*.
The five lower energy data points are from the EGRET source 3EG J1746-2851.
The
theoretical curves are:
(blue) {\bf solid}:
$\gamma$-ray spectrum from the entire CND 
after $\chi^2$ fitting to the $X$ parameter {\it to the HESS data only};
(red) {\bf short dashed}:
$\gamma$-ray spectrum from the entire CND 
after $\chi^2$ fitting of the $X$ parameter to the HESS data
{\it with the additional constraint that the EGRET upper limit is obeyed};
(purple) {\bf long dashed}: 
$\gamma$-ray spectrum from the wind region clumps assuming the same $X$ value;
(yellow) {\bf dot-dash}: $\gamma$-ray
spectrum for the case of a single zone (with $B = 0.35$ mG
and $n_H \simeq 2.3 \times 10^5$ cm$^{-3}$) model and
a single, power-law proton population;
(green) {\bf dot-dot-dash}: projected
GLAST sensitivity for a 5 $\sigma$ detection of a point-source with a $\propto E^2$ spectrum
(see the GLAST website: http://www-glast.stanford.edu).
{\bf b.} ({\bf right}) Observed radio fluxes from the CND and wind regions together
with theoretical predictions for radio emission.
The theoretical radio curves are: 
(blue) {\bf solid}: radio emission from the {\bf CND} region
for the case that the theoretical CND $\gamma$-ray emission is optimized in $X$ to the HESS data;
(purple) {\bf long dashed}: radio emission from the {\bf wind} region
assuming the same $X$ value;
(red) {\bf short dashed}:
radio emission from the {\bf wind} region
for the case that the theoretical CND $\gamma$-ray emission is optimized in $X$ to the HESS data
{\it with the additional constraint that the EGRET upper limit is obeyed};
(green) {\bf dotted} curve:
simple power law fit to 843, 1384, and 2368 MHz flux
points.}
\label{fig:CNDRadioSpectrum}
\end{figure}

We find a minimum reduced
$\chi^2$ value of 1.2 (for 32 degrees of freedom) at $X = 1.69$.
Extension of individual clump spectra below TeV
using the power-law assumption
then allows us to also
make a prediction for the synchrotron radio emission due to secondary leptons
in each CND clump and, using the same $X$ value, in each cell of the wind region too
(our procedure here
takes into account cooling by the ionization, bremsstrahlung and synchrotron
processes to arrive at
the steady-state electron and positron distributions within each clump; 
see \citet{Crocker2007a} for details).
The same extension -- to still lower proton energies --
allows us to also predict the lower-energy $\gamma$-ray spectrum from the CND
(and also the wind region).
The total $\gamma$-ray and synchrotron radio
emission curves for all cells in the CND
are shown as the solid (blue)
curves in Figures \ref{fig:CNDRadioSpectrum} (a) and (b), respectively.
Gamma-ray and radio emission by cells in the wind region is shown
by the long-dashed (purple) curves in the same figures.
As may be seen, the $\gamma$-ray normalized secondary synchrotron radio flux
from the CND directly 
accounts for $\sim$ 10 \% of the observed GHz radio emission but with a rather flatter spectrum
than the observation data suggest.
In contrast,
the wind region is modeled to produce a radio spectrum consistent with observation
but with
around twice the observed normalization.
Given that this determination involves absolutely no fine tuning and is predicated on
a fit of theoretical $\gamma$-ray emission {\it from a separate region} (i.e., the CND clumps) 
to the TeV HESS data, 
this is a remarkable level of agreement and we conclude that our 
secondary emission model can self-consistently account for both the $\gamma$-ray emission
seen from the direction of the GC and the total radio emission from the wind and CND regions
(i.e., a $2 \times 1'$ region centered in Sgr A*).

In Figure \ref{fig:CNDRadioSpectrum} (a) one notes that
the CND $\gamma$-ray emission for the scenario above
is roughly consistent with the level of emission seen from the
EGRET source 3EG~J1746-2851 \citep{Mayer-Hasselwander1998}
at the two lowest energy data points around 70 MeV.
We do not claim, however, that the CND emission explains the origin of
the observed \grs .
In fact, recent studies have shown that the EGRET source excludes the GC
at the $99.9\%$ confidence level \citep{Hooper2002,Pohl2005}.
On the other hand, the position of the {\it predicted}
CND emission was certainly inside the wide field of view
of the EGRET instrument's GC pointings.
The spectrum of 3EG~J1746-2851 serves, then,
as an upper limit to the allowed emission from the CND.
The (blue) solid curve would seem to be just excluded.
A fit to the HESS data with the additional constraint that the extrapolated $\gamma$-ray emission
obey the constraint that it be unobserved by EGRET
(which we translate to the requirement that the predicted low energy gamma ray curve 
pass 2$\sigma$ below the most constraining EGRET datum near 70 MeV)
requires $X \simeq 0.8$ and the radio spectrum obtained in this case, 
shown as the (red) short-dashed curve in 
\ref{fig:CNDRadioSpectrum} (b), is statistically compatible with the radio data.
The reduced $\chi^2$ for this case is rather bad, however: 2.2 for 32 degrees of freedom.
The CND $\gamma$-ray spectrum for this case is shown as the (red) short-dashed curve in 
\ref{fig:CNDRadioSpectrum} (a).

Also shown in Figures \ref{fig:CNDRadioSpectrum} (a) and (b)
are (yellow) dot-dash curves that show, respectively,
the predicted gamma-ray and radio emission for a single zone model
that assumes a single power-law $p$ population ($\sim E^{-2.3}$)
tuned to match the HESS emission.
With such a model one notes that, as first determined by
\citet{Crocker2005}, the low energy $\gamma$-ray curve passes well below the EGRET
points. 
But the model predicts far too little synchrotron when compared with
our full calculations that sum emission over each clump of the CND and wind regions with its own
$p$ spectrum, magnetic field and ambient hydrogen number density.
This underestimation probably arises from the assumption of a `mean' 
$B$ field within the single zone model that implictly neglects the (i) 
relative accumulation of protons into clumps of higher magnetic field strength
(where they are more strongly trapped than elsewhere) and (ii) the greater
synchrotron emissivity of secondaries created in such clumps.


Finally,
we have reviewed data available from observations of the GC and the CND and wind regions at other wavelengths
to see whether these
offer any further constraint on our model.

Given the steady-state positron production rate in our scenario we 
predict a 511 keV $\gamma$-ray production rate from 
electron-positron annihilation of $\sim 6 \times 10^{46}$ yr$^{-1}$, well
inside the limit from INTEGRAL observations ($\sim 10^{50}$ yr$^{-1}$ out to an angular radius of 8$^\circ$: 
\cite{Knodlseder2003}

Another possible constraint is offered by hard X-ray/soft $\gamma$-ray observations. 
Our model predicts diffuse secondary emission at energies of
10$^{4-5}$ eV
due to both the bremsstrahlung
and, dominantly, synchrotron processes. 
We have modeled such emission
to compare it against the spectrum, at a comparable energy, of the INTEGRAL/IBIS source
IGRJ17456-2901 \citep{Belanger2004}; we find ourselves well below the level of flux from this source.


\section{Conclusions}
\label{sect:Conclusions}

There are several major conclusions we can draw from this study:

\noindent
(1) The pre-existing data were not sufficient for a tight constraint on
our predictions. New data were acquired for this project. The compilation
of data demonstrates that the black hole-induced hadronic model for the
TeV gamma rays 
introduced in \citet{Ballantyne2007}
is certainly consistent with the broadband emission from
this process at other wavelengths.

\noindent
(2) In fact, more strongly, we have also shown that, {\it without fine-tuning, 
the hadronic process may actually explain essentially all the diffuse 
radio emission observed from the inner few pc of the Galaxy}.



\noindent
(3) The fact that in this model the protons energized by the black hole
and ejected into the ISM accumulate in the magnetic fields of the CND, 
provides a spatial
definition for where the secondaries will be active unlike, say, the
primary lepton model, in which the electrons may be accelerated wherever
shock fragments form in an expanding shell.



\noindent
(4) The propagation calculation of \citet{Ballantyne2007} 
showed that diffusion processing by the GC magnetic field leads 
to a considerable spectral steepening of the injection spectrum, $\sim E^{1.5}$
requiring in turn, that the  spectrum of protons injected 
close to Sgr A* must be very flat, $\sim~E^{-0.75}$, 
in order to supply the requisite $\sim~E^{-2.25}$ spectrum of $\gamma$-rays.
This is much flatter than the $E^{-(2.1 \to 2.4)}$ spectrum expected from 
first-order Fermi acceleration. 
Such a  hard spectrum might be created via 
the (second-order Fermi) stochastic acceleration mechanism. 
For instance, as determined by \citet{Liu2006} stochastic acceleration in a 
magnetically-dominated funnel close to the black hole 
could accelerate protons into a distribution as flat as $E^{-1}$, 
approaching the requisite hardness. 
The stochastic acceleration mechanism
investigated by \citet{Becker2006} may
also be able to produce the required spectrum. 
The question of whether a stochastic acceleration can be made to work in this context 
will be addressed elsewhere by the current 
authors.

\noindent
(5) Finally, the hadronic scenario we have explored here 
may be a test bed of what is actually happening near the base of
relativistic jets in more powerful sources, such as AGNs and quasars.
There we would have a more difficult time discerning the various
processes, because the environment is dense and chaotic. But in Sgr A*,
the environment is much more sedate, with lower density, and less
activity. Sgr A* does not itself produce relativistic jets that we
can see. But that may simply be a consequence of the relative 
weakness of this process in this particular source. It may be that
something like the proton injection/acceleration scenario 
(with
subsequent propagation out to relatively large scales before final interaction)
posited in our earlier paper -- and whose phenomenological consequences we have explored here --
is happening on a much bigger scale in the more
powerful AGNs.

\acknowledgments

The authors thank Todor Stanev for producing the simulated
secondary particle spectra from pp collisions exploited in this paper
and Anne Green for providing the SUMSS GC 843 MHz data in numerical form.
DRB is supported by the University of Arizona Theoretical Astrophysics
Program Prize Postdoctoral Fellowship.
RMC gratefully acknowledges advice and assistance from Ray Protheroe.
RMC is supported at the University of Adelaide by Ray Protheroe
and Ron Ekers'
Australian Research Council's Discovery funding scheme grant
(project number
DP0559991). 
This work was funded, in part, at the University of Arizona
by NSF grant AST-0402502.
The work has made use
of NASA's Astrophysics Data System Abstract Service.
FM is
grateful to the University of Melbourne for its support (through a Sir
Thomas Lyle Fellowship and a Miegunyah Fellowship).

{}

\clearpage

\begin{thebibliography}{}


\bibitem[\protect\citeauthoryear{Aharonian \etal}{2004}]{Aharonian2004}
  Aharonian, F.A. \etal, 2004, \aap, 425, 13

\bibitem[\protect\citeauthoryear{Aharonian \etal}{2006a}]{Aharonian2006}
  Aharonian, F.A. \etal, 2006a, Nature, 439, 695

\bibitem[\protect\citeauthoryear{Aharonian \etal}{2006b}]{Aharonian2006b}
  Aharonian, F.A. \etal, 2006b, \prl, 97, 221102



\bibitem[\protect\citeauthoryear{Baganoff \etal}{2003}]{Baganoff2003}
  Baganoff, F.K. \etal, 2003, \apj, 591, 891

\bibitem[Ballantyne \etal (2007)]{Ballantyne2007} Ballantyne, D.~R.,
Melia, F., Liu, S., \& Crocker, R.~M.\ 2007, \apjl, 657, L13

\bibitem[\protect\citeauthoryear{Becker \etal}{2006}]{Becker2006} Becker,
  P.A., Le, T. \& Dermer, C.D., 2006, \apj, 647, 539

\bibitem[Belanger \etal (2004)]{Belanger2004} Belanger, G.,
\etal \ 2004, \apjl, 601, L163


\bibitem[Brogan \etal (2003)]{Brogan2003} Brogan, C., \etal, 2003,
Astronomische Nachrichten Supplement, 324, 17

\bibitem[Brown et al.(1981)]{Brown1981} Brown, R.~L., Johnston,
K.~J., \& Lo, K.~Y.\ 1981, \apj, 250, 155

\bibitem[\protect\citeauthoryear{Crocker \etal}{2005}]{Crocker2005}
  Crocker, R.M., Fatuzzo, M., Jokipii, J.R., Melia, F. \& Volkas,
  R.R. 2005, \apj, 622, 892

\bibitem[Crocker \etal(2007a)]{Crocker2007a}Crocker, R.M., \etal, 2007a \apj \ {\it in press},
(astro-ph/0702045).


\bibitem[Crutcher(1999)]{Crutcher1999} {Crutcher}, R.~M.\ 1999, \apj, 520, 706



\bibitem[Ekers et al.(1983)]{Ekers1983} Ekers, R.~D., van Gorkom, 
J.~H., Schwarz, U.~J., \& Goss, W.~M.\ 1983, \aap, 122, 143 

\bibitem[Fatuzzo \& Melia(2003)]{Fatuzzo2003} Fatuzzo, M., \& 
Melia, F.\ 2003, \apj, 596, 1035 


\bibitem[Gopal-Krishna \& Swarup(1976)]{Gopal-Krishna1976} Gopal-Krishna, 
\& Swarup, G.\ 1976, \aplett, 17, 45 



\bibitem[\protect\citeauthoryear{Hooper \& Dingus}{2002}]{Hooper2002}
  Hooper, D. \& Dingus, B. 2002, in the proceedings of the 34th COSPAR
  Scientific Assembly (astro-ph/0212509)


\bibitem[Jones \etal (2007)]{Jones2007} Jones, D.I., \etal , in prep., 2007.


\bibitem[Kn{\"o}dlseder et al.(2003)]{Knodlseder2003} Kn{\"o}dlseder, 
J., et al.\ 2003, \aap, 411, L457 



\bibitem[LaRosa \etal (2000)]{LaRosa2000} LaRosa, T., \etal, AJ, 119, 107, 2000.



\bibitem[\protect\citeauthoryear{Liu \etal}{2006}]{Liu2006} Liu, S.,
  Melia, F., Petrosian, V. \& Fatuzzo, M. 2006, \apj, 647, 1099

\bibitem[Mayer-Hasselwander \etal(1998)]{Mayer-Hasselwander1998}
Mayer-Hasselwander, H.~A., \etal\ 1998, \aap, 335, 161

\bibitem[\protect\citeauthoryear{Melia}{2007}]{Melia2007} Melia, F., 2007,
  The Galactic Supermassive Black Hole, Princeton University Press



\bibitem[Pedlar \etal (1989)]{Pedlar1989} Pedlar, A.,
Anantharamaiah, K.~R., Ekers, R.~D., Goss, W.~M., van Gorkom, J.~H.,
Schwarz, U.~J., \& Zhao, J.-H.\ 1989, \apj, 342, 769

\bibitem[\protect\citeauthoryear{Pohl}{2005}]{Pohl2005} Pohl, M. 2005,
  \apj, 626, 174


\bibitem[\protect\citeauthoryear{Rockefeller \etal}{2004}]{Rockefeller2004}
  Rockefeller, G., Fryer, C. L., Melia, F. \& Warren, M. S. 2004,
  \apj, 604, 662

\bibitem[Salter \etal (1988)]{Salter1988} Salter, C.~J., Sinha,
R.~P., Stobie, E.~B., Kerr, F.~J., \& Hobbs, R.~W.\ 1988, \mnras, 232, 407


\bibitem[Sofue \etal (1986)]{Sofue1986} Sofue, Y., Inoue, M.,
Handa, T., Tsuboi, M., Hirabayashi, H., Morimoto, M., \& Akabane, K.\ 1986,
\pasj, 38, 475

\bibitem[Tsuboi \etal (1988)]{Tsuboi1988} Tsuboi, M., Handa, T.,
Inoue, M., Ukita, N., \& Takano, T.\ 1988, \pasj, 40, 665




\end{thebibliography}
\end{document}